\documentclass{DISproc}

\def\d{\mathrm{d}}
\begin{document}
\title{An AdS/QCD holographic wavefunction for the $\rho$ meson}

\author{{\slshape Jeff Forshaw$^1$, Ruben Sandapen$^2$}\\[1ex]
$^1$University of Manchester, Oxford Road, Manchester M13 9PL, UK.\\
$^2$Universit\'e de Moncton, Moncton, N-B, E1A 3E9, Canada.}

\contribID{265}

\doi  

\maketitle

\begin{abstract}
We use an AdS/QCD holographic wavefunction to generate predictions for the rate of diffractive $\rho$-meson
electroproduction that are in reasonable agreement with data collected
at the HERA electron-proton collider. 
\end{abstract}

\section{Introduction}
In the dipole model of high-energy scattering \cite{Nikolaev:1990ja,Nikolaev:1991et,Mueller:1993rr,Mueller:1994jq}, the scattering amplitude for diffractive $\rho$ meson production is a convolution of the photon
and vector meson $q\bar{q}$ light-front wavefunctions with the total cross-section
to scatter a $q\bar{q}$ dipole off a proton. QED is used to determine
the photon wavefunction and the dipole cross-section can be extracted from the precise data on the deep-inelastic structure
function $F_2$ \cite{Soyez:2007kg,Forshaw:2004vv}. This formalism can then be used to predict rates for vector meson production and diffractive DIS \cite{Forshaw:2003ki,Forshaw:2006np} or to
to extract information on the $\rho$ meson wavefunction using the HERA data on diffractive $\rho$ production \cite{Forshaw:2010py,Forshaw:2011yj}. Here we use it to predict the cross-sections for diffractive $\rho$ production using an AdS/QCD holographic wavefunction proposed by Brodsky and de T\'eramond \cite{deTeramond:2008ht}. We also compute the second moment of the twist-$2$ distribution amplitude and find it to be in agreement with Sum Rules and lattice predictions.

\section{The AdS/QCD holographic wavefunction}
In a semiclassical approximation to light-front QCD  the meson 
wavefunction can be written in the
following factorized form \cite{deTeramond:2008ht}
\begin{equation}
\phi(x,\zeta, \varphi)=\frac{\Phi(\zeta)}{\sqrt{2\pi \zeta}} f(x) \mathrm{e}^{i L \varphi} 
\label{factorized-lc}
\end{equation}
where $L$ is the orbital quantum number and $\zeta=\sqrt{x(1-x)} b$ ($x$
is the light-front longitudinal momentum fraction of the quark
and $b$ the quark-antiquark transverse separation). The function $\Phi(\zeta)$ satisfies a
Schr\"odinger-like wave equation
\begin{equation}
\left(-\frac{\d^2}{\d \zeta^2} - \frac{1-4L^2}{4 \zeta^2} + U(\zeta) \right) \Phi (\zeta)=M^2 \Phi (\zeta) \;,
\label{LFeigenvalue}
\end{equation}
where $U(\zeta)$ is the confining potential defined at equal light-front time. After identifying $\zeta$ with
the co-ordinate in the fifth dimension in AdS space, Eq.~\eqref{LFeigenvalue}
describes the propagation of spin-$J$ string modes, in which
case $U(\zeta)$ is determined by the choice for the dilaton
field. We use here the soft-wall
model~\cite{Karch:2006pv}, in which 
\begin{equation}
 U(\zeta)=\kappa^4 \zeta^2 + 2\kappa^2(J-1) \;.
\label{quadratic-dilaton}
\end{equation}
The eigenvalues of Eq.~\eqref{LFeigenvalue} are then given as
\begin{equation}
 M^2=4\kappa^2(n+J/2+L/2) \; , 
\label{mass-spectrum}
\end{equation}
so that the parameter $\kappa$ can then be fixed as the best fit value to the Regge slope for vector mesons. Here we use $\kappa=0.55$ GeV.  After solving Eq.~\eqref{LFeigenvalue} with $L=0$ and $S=1$ to obtain $\Phi(\zeta)$, it remains to
specify the function $f(x)$ in equation~\eqref{factorized-lc}. This is done by
comparing the expressions for the pion EM form factor
obtained in the light-front formalism and in AdS
space \cite{Brodsky:2007hb}.
After accounting for non zero quark masses \cite{Brodsky:2008pg}, the final form of the AdS/QCD wavefunction is \cite{Forshaw:2012im}
\begin{equation}
 \phi(x,\zeta)= N \frac{\kappa}{\sqrt{\pi}}\sqrt{x(1-x)} \exp \left(-\frac{\kappa^2 \zeta^2}{2}\right) \exp\left(-\frac{m_f^2}{2\kappa^2 x (1-x)} \right)~,
\label{lcwf-massive-quarks}
\end{equation}
where $N$ is fixed so that 
\begin{equation}
\int {\d}^2{\mathbf{b}} \; \d x \; |\phi(x,\zeta)|^2 = 1~.
\label{eq:norm}
\end{equation}

The meson's light-front wavefunctions can be written in terms of the
AdS/QCD wavefunction $\phi(x,\zeta)$ \cite{Forshaw:2011yj}. For longitudinally
polarized mesons:
\begin{equation}
\Psi^{L}_{h,\bar{h}}(b,x) = \frac{1}{2\sqrt{2}}
\delta_{h,-\bar{h}} 
\left( 1 +  \frac{ m_{f}^{2} -  \nabla^{2}}{M_{\rho}^2\; x(1-x)}\right) \phi(x,\zeta) ~,
\label{nnpz_L}
\end{equation}
where $\nabla^2 \equiv \frac{1}{b} \partial_b + \partial^2_b$ and
$h$ ($\bar{h}$) are the helicities of the quark (anti-quark). The
imposition of current conservation implies that this can be replaced by
\begin{equation}
\Psi^{L}_{h,\bar{h}}(b,x) = \frac{1}{\sqrt{2}}
\delta_{h,-\bar{h}} \; \phi(x,\zeta) ~.
\label{nnpz_L1}
\end{equation}
We choose to normalize $\phi(x,\zeta)$ using
\begin{equation}
\sum_{h,\bar{h}}\int \d^{2}{\mathbf{b}} \, \d x  \,
|\Psi^{L}_{h,\bar{h}}(b, x)|^{2} = 1 ~
\label{normalisation}
\end{equation}
using either Eq.~\eqref{nnpz_L} or Eq.~\eqref{nnpz_L1} and refering to them as Method B or Method A respectively. Note that Method A implies that Eq.~\eqref{eq:norm}  is satisfied exactly whereas Method B is equivalent to assuming that the integral in Eq.~\eqref{eq:norm} is a little larger than unity. 

For transversely polarized mesons:
\begin{equation}
\Psi^{T=\pm}_{h,\bar{h}}(b, x) = \pm [i e^{\pm i\theta} 
( x \delta_{h\pm,\bar{h}\mp} - (1-x) \delta_{h\mp,\bar{h}\pm}) 
\partial_{b}+ m_{f}\delta_{h\pm,\bar{h}\pm}] \frac{\phi(x,\zeta)}{2x(1-x)}~,
\label{nnpz_T}
\end{equation}
where $be^{i\theta}$ is the complex form of the transverse separation,
$\mathbf{b}$.

\section{Comparing to data, sum rules and the lattice}
\begin{figure}[htb]
  \centering
  \includegraphics[width=0.4\textwidth]{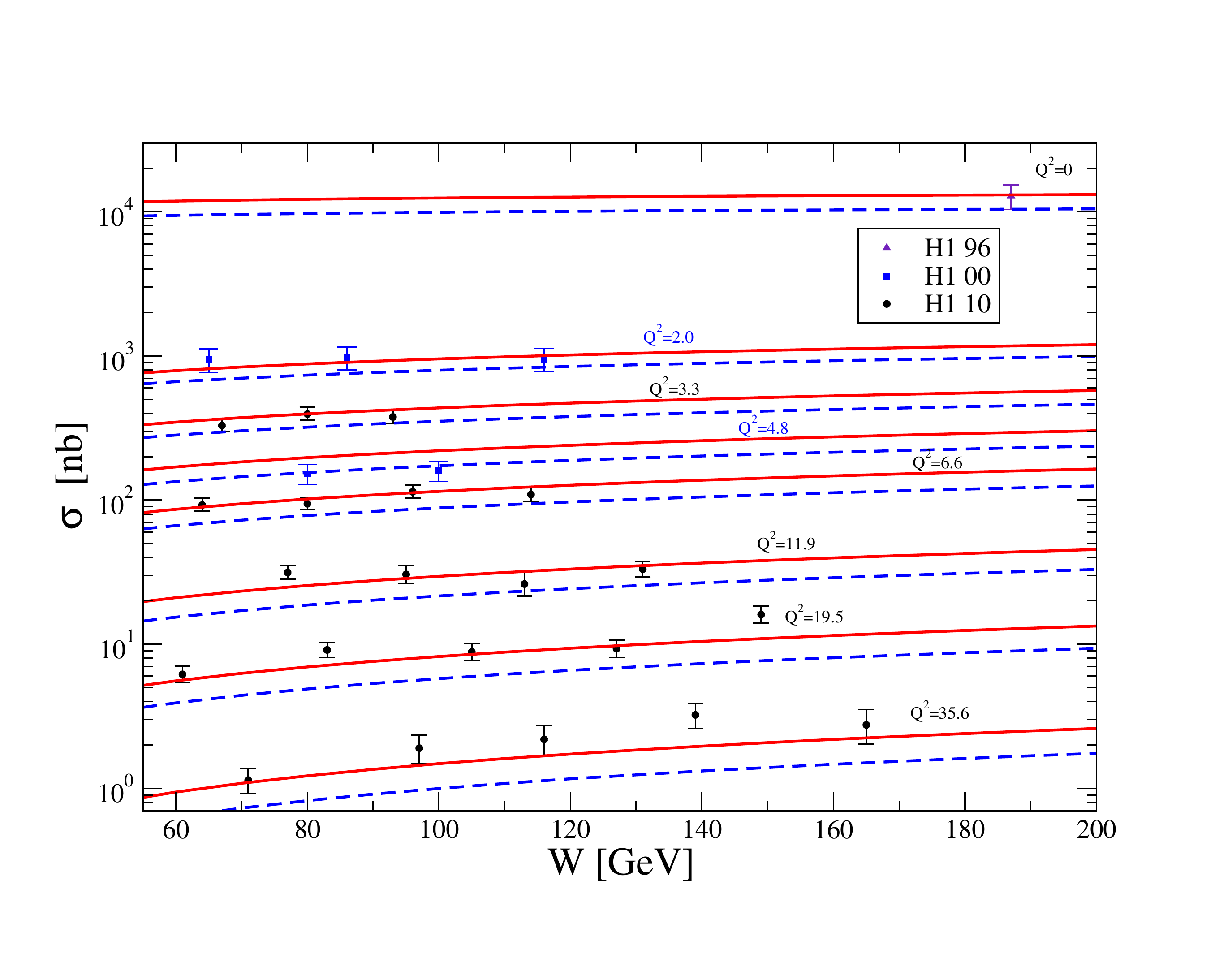} \includegraphics[width=0.4\textwidth]{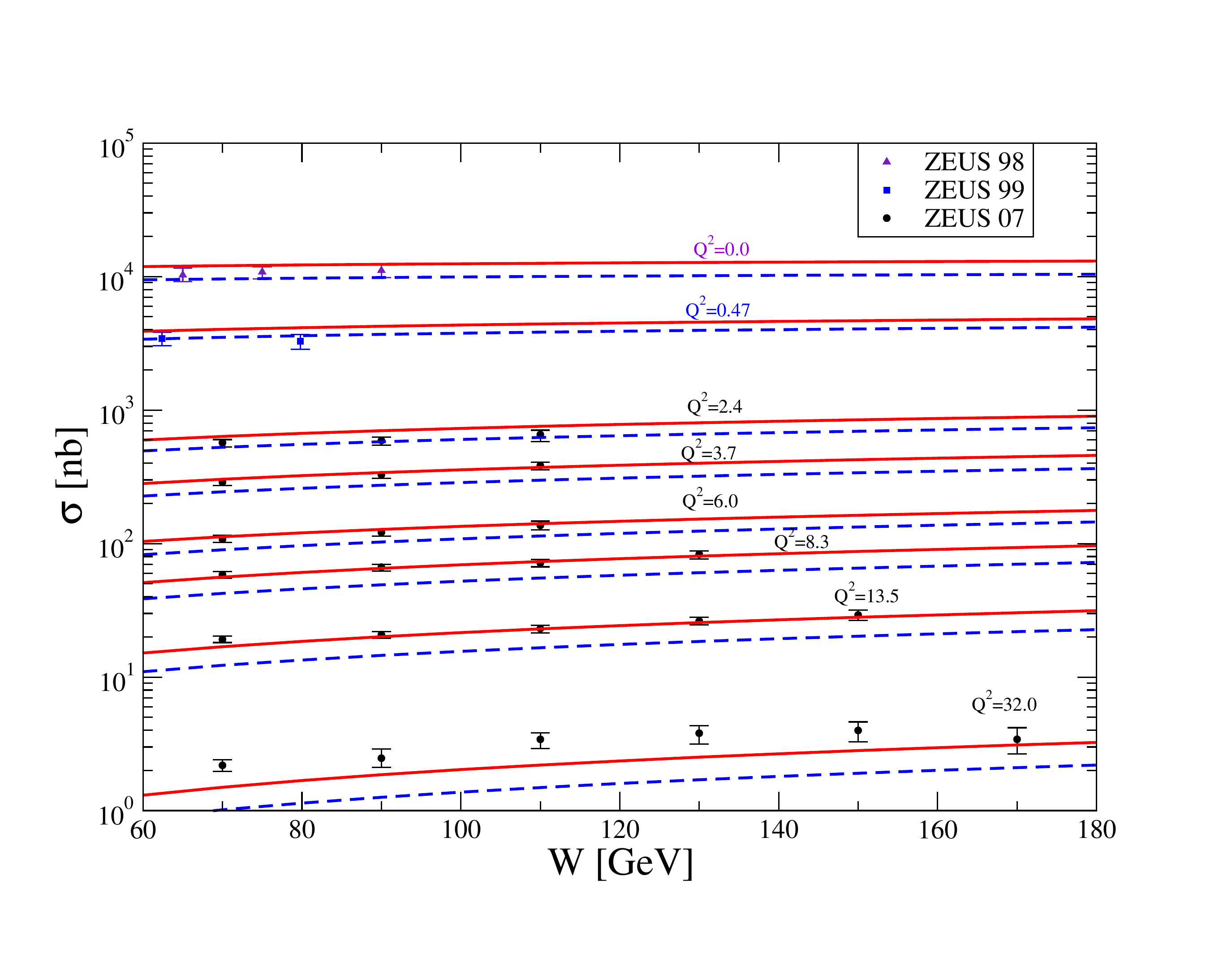} \includegraphics[width=0.4\textwidth]{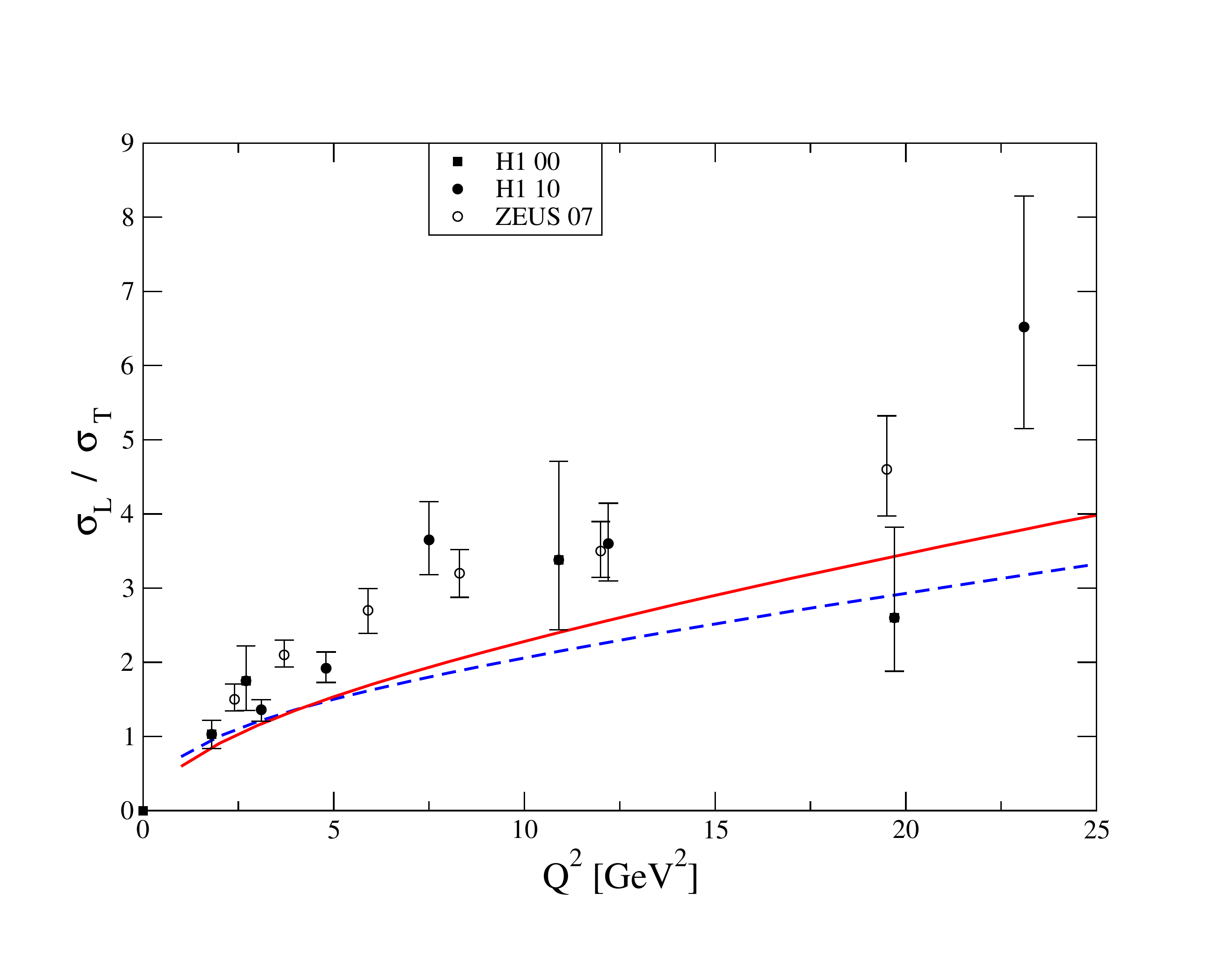}
  \caption{Comparison to the HERA data
  \cite{Chekanov:2007zr,Collaboration:2009xp}. Solid red curve is for
  Method B and the dashed blue curve is for Method A.}
  \label{Fig:data}
\end{figure}

Our predictions for the total cross-section and the ratio of longitudinal to transverse cross-section are compared to the HERA data in Fig.~\ref{Fig:data}. As can be seen, the agreement is quite good given that our predictions do not contain any free parameters. The disagreement at high $Q^2$ is expected since this is the region where perturbative evolution of the
wavefunction will be relevant and the AdS/QCD wavefunction we use is
clearly not able to describe that.

We also compute the second moment of the corresponding twist-$2$ Distribution Amplitude and find our predictions to be in agreement with those made using Sum Rules and lattice QCD. We obtain a value of $0.217$ for Method A and $0.228$ for Method B, which is to be compared with the
Sum Rule result of $0.24 \pm 0.02$ at $\mu = 3$~GeV \cite{Ball:2007zt} and the lattice
result of $0.24\pm 0.04$ at $\mu = 2$~GeV \cite{Boyle:2008nj}.  The AdS/QCD
wavefunction neglects the perturbatively known evolution with the
scale $\mu$ and should be viewed as a parametrization of the DA at
some low scale $\mu \sim 1$ GeV. Viewed as such, the agreement is
good. 

\section{Acknowledgements}
R.S thanks the organisers for a very pleasant workshop and  the Facult\'e des Sciences of the Universit\'e de Moncton as well as the Facult\'e des \'Etudes Superieures et de la Recherche (FESR) of the Universit\'e de Moncton for financial support. 

\bibliographystyle{DISproc}
\bibliography{sandapen_ruben}

\providecommand{\href}[2]{#2}\begingroup\raggedright\begin{thebibliography}{10}

\bibitem{Nikolaev:1990ja}
N.~N. Nikolaev and B.~G. Zakharov.
\newblock
\href{http://dx.doi.org/10.1007/BF01483577}{Z. Phys. {\bfseries C49} (1991)
  607--618}.

\bibitem{Nikolaev:1991et}
N.~N. Nikolaev and B.~G. Zakharov.
\newblock
\href{http://dx.doi.org/10.1007/BF01597573}{Z. Phys. {\bfseries C53} (1992)
  331--346}.

\bibitem{Mueller:1993rr}
A.~H. Mueller.
\newblock
\href{http://dx.doi.org/10.1016/0550-3213(94)90116-3}{Nucl. Phys. {\bfseries
  B415} (1994) 373--385}.

\bibitem{Mueller:1994jq}
A.~H. Mueller and B.~Patel.
\newblock \href{http://dx.doi.org/10.1016/0550-3213(94)90284-4}{Nucl. Phys.
  {\bfseries B425} (1994) 471--488},
\href{http://arxiv.org/abs/hep-ph/9403256}{{\ttfamily arXiv:hep-ph/9403256}}.

\bibitem{Soyez:2007kg}
G.~Soyez.
\newblock \href{http://dx.doi.org/10.1016/j.physletb.2007.07.076}{Phys. Lett.
  {\bfseries B655} (2007) 32--38},
\href{http://arxiv.org/abs/0705.3672}{{\ttfamily arXiv:0705.3672 [hep-ph]}}.

\bibitem{Forshaw:2004vv}
J.~R. Forshaw and G.~Shaw.
\newblock \href{http://dx.doi.org/10.1088/1126-6708/2004/12/052}{JHEP
  {\bfseries 12} (2004) 052},
\href{http://arxiv.org/abs/hep-ph/0411337}{{\ttfamily arXiv:hep-ph/0411337}}.

\bibitem{Forshaw:2003ki}
J.~R. Forshaw, R.~Sandapen, and G.~Shaw.
\newblock \href{http://dx.doi.org/10.1103/PhysRevD.69.094013}{Phys. Rev.
  {\bfseries D69} (2004) 094013},
\href{http://arxiv.org/abs/hep-ph/0312172}{{\ttfamily arXiv:hep-ph/0312172}}.

\bibitem{Forshaw:2006np}
J.~R. Forshaw, R.~Sandapen, and G.~Shaw.
\newblock JHEP {\bfseries 11} (2006) 025,
\href{http://arxiv.org/abs/hep-ph/0608161}{{\ttfamily arXiv:hep-ph/0608161}}.

\bibitem{Forshaw:2010py}
J.~R. Forshaw and R.~Sandapen.
\newblock \href{http://dx.doi.org/10.1007/JHEP11(2010)037}{JHEP {\bfseries 11}
  (2010) 037},
\href{http://arxiv.org/abs/1007.1990}{{\ttfamily arXiv:1007.1990 [hep-ph]}}.

\bibitem{Forshaw:2011yj}
J.~R. Forshaw and R.~Sandapen.
\newblock \href{http://dx.doi.org/10.1007/JHEP10(2011)093}{JHEP {\bfseries
  1110} (2011) 093}, \href{http://arxiv.org/abs/1104.4753}{{\ttfamily
  arXiv:1104.4753 [hep-ph]}}.

\bibitem{deTeramond:2008ht}
G.~F. de~Teramond and S.~J. Brodsky.
\newblock
  \href{http://dx.doi.org/10.1103/PhysRevLett.102.081601}{Phys.Rev.Lett.
  {\bfseries 102} (2009) 081601},
  \href{http://arxiv.org/abs/0809.4899}{{\ttfamily arXiv:0809.4899 [hep-ph]}}.

\bibitem{Karch:2006pv}
A.~Karch, E.~Katz, D.~T. Son, and M.~A. Stephanov.
\newblock \href{http://dx.doi.org/10.1103/PhysRevD.74.015005}{Phys.Rev.
  {\bfseries D74} (2006) 015005},
  \href{http://arxiv.org/abs/hep-ph/0602229}{{\ttfamily arXiv:hep-ph/0602229
  [hep-ph]}}.

\bibitem{Brodsky:2007hb}
S.~J. Brodsky and G.~F. de~Teramond.
\newblock \href{http://dx.doi.org/10.1103/PhysRevD.77.056007}{Phys.Rev.
  {\bfseries D77} (2008) 056007},
  \href{http://arxiv.org/abs/0707.3859}{{\ttfamily arXiv:0707.3859 [hep-ph]}}.

\bibitem{Brodsky:2008pg}
S.~J. Brodsky and G.~F. de~Teramond.
\newblock
\href{http://arxiv.org/abs/0802.0514}{{\ttfamily arXiv:0802.0514 [hep-ph]}}.

\bibitem{Forshaw:2012im}
J.~Forshaw and R.~Sandapen.
\newblock
\href{http://arxiv.org/abs/1203.6088}{{\ttfamily arXiv:1203.6088 [hep-ph]}}.

\bibitem{Ball:2007zt}
P.~Ball, V.~M. Braun, and A.~Lenz.
\newblock JHEP {\bfseries 08} (2007) 090,
\href{http://arxiv.org/abs/0707.1201}{{\ttfamily arXiv:0707.1201 [hep-ph]}}.

\bibitem{Boyle:2008nj}
P.~A. Boyle {\em et~al.}
\newblock PoS {\bfseries LATTICE2008} (2008) 165,
\href{http://arxiv.org/abs/0810.1669}{{\ttfamily arXiv:0810.1669 [hep-lat]}}.

\bibitem{Chekanov:2007zr}
S.~Chekanov {\em et~al.}
\newblock \href{http://dx.doi.org/10.1186/1754-0410-1-6}{PMC Phys. {\bfseries
  A1} (2007) 6},
\href{http://arxiv.org/abs/0708.1478}{{\ttfamily arXiv:0708.1478 [hep-ex]}}.

\bibitem{Collaboration:2009xp}
{The H1 Collaboration}.
\newblock \href{http://dx.doi.org/10.1007/JHEP05(2010)032}{JHEP {\bfseries 05}
  (2010) 032},
\href{http://arxiv.org/abs/0910.5831}{{\ttfamily arXiv:0910.5831 [hep-ex]}}.

\end{thebibliography}\endgroup

\end{document}